\documentclass[]{aa}
\usepackage{graphicx, natbib, longtable, lscape, amssymb}
\usepackage[english]{babel}
\usepackage{lscape}

\newcommand{\gppr}{\stackrel{>}{\scriptstyle \sim}}
\newcommand{\gappr}{\raisebox{-0.4ex}{$\gppr$}}
\newcommand{\lppr}{\stackrel{<}{\scriptstyle \sim}}
\newcommand{\lappr}{\raisebox{-0.4ex}{$\lppr$}}

\newcommand{\Porb}{\mbox{$P_\mathrm{orb}$}}

\newcommand{\Mwd}{\mbox{$M_\mathrm{wd}$}}

\newcommand{\Msun}{\mbox{$\mathrm{M}_{\odot}$}}

\newcommand{\Rsun}{\mbox{$R_{\odot}$}}

\begin{document}

\title{Origin of apparent period variations in eclipsing post-common-envelope binaries} 
\titlerunning{Origin of apparent period variations in PCEBs}
\author{M. Zorotovic\inst{1}, M.R. Schreiber\inst{1,2}}
\authorrunning{Zorotovic \& Schreiber}
\institute{Departamento de F\'isica y Astronom\'ia, Facultad de Ciencias, Universidad de Valpara\'iso, Valpara\'iso, Chile \\
\email{mzorotovic@dfa.uv.cl}
\and{Millennium Nucleus ``Protoplanetary Disks in ALMA Early Science",
  Universidad de Valpara\'iso, Valpara\'iso, Chile}
}
\offprints{M. Zorotovic}

\date{Received 3 September 2012 / Accepted 6 November 2012}

\abstract{Apparent period variations detected in several eclipsing, close-compact binaries are frequently interpreted as being caused by circumbinary giant planets. This interpretation raises the question of the origin of the potential planets that must have either formed in the primordial circumbinary disk, together with the host binary star, and survived its evolution into a close-compact binary or formed in a post-common-envelope circumbinary disk that remained bound to the post-common-envelope binary (PCEB).} 
{Here we combine current knowledge of planet formation and the statistics of giant planets around primordial and evolved binary stars with the theory of close-compact binary star evolution aiming to derive new constraints on possible formation scenarios.}
{We compiled a comprehensive list of observed eclipsing PCEBs, estimated the fraction of systems showing apparent period variations, reconstructed the evolutionary history of the PCEBs, and performed binary population models of PCEBs to characterize their main sequence binary progenitors. We reviewed the currently available constraints on the fraction of PCEB progenitors that host circumbinary giant planets.}
{We find that the progenitors of PCEBs are very unlikely to be frequent hosts of giant planets ($\lappr\,10$ per cent), while the frequency of PCEBs with observed apparent period variations is very high ($\sim\,90$ per cent). }
{The variations in eclipse timings measured in eclipsing PCEBs are probably not caused by first-generation planets that survived common-envelope evolution. The remaining options for explaining the observed period variations are second-generation planet formation or perhaps variations in the shape of a magnetically active secondary star. We suggest observational tests for both options.}

\maketitle

\section{Introduction}\label{sec:intro}

The discovery of extrasolar planets (exoplanets) in the nineties and the subsequent detection of more than 700 exoplanets until now \citep{schneideretal11-1} has taken us much closer toward answering fundamental questions, such as how planets form from circumstellar material and what the fate of planetary systems is when their host stars evolve into giant stars and compact objects. Especially that exoplanets have been detected in rather unexpected exotic environments might go further pave our way towards solving these important questions. 

Pulsar timings have led to detecting the first confirmed exoplanet orbiting the pulsar PSR\,B1257+12 \citep{wolszczan+frail92-1,wolszczan94-1}, and continued monitoring of the system has revealed the existence of a second and a third planet \citep{wolszczan94-1}, even the possibility of a fourth one \citep{wolszczan97-1}. The first exoplanet around a solar-type star, a Jupiter-mass companion to the star 51 Pegasi, was identified and confirmed thanks to radial velocity variations \citep{mayor+queloz95-1}. This finding generated large radial velocity surveys for exoplanets, and meanwhile more than 600 planets around main sequence (MS) stars have been identified using this method. Circumbinary exoplanets around binary stars composed of two MS stars (MS+MS binaries) were predicted long ago, but only recently have the first six detections been possible \citep{doyleetal11-1,welshetal12-1,oroszetal12-1,oroszetal12-2}.   

Among the perhaps most unexpected potential planet detections are those derived from apparent period variations (or eclipse time variations), frequently measured in eclipsing, close-compact binary stars containing a white dwarf (WD) or hot subdwarf B (sdB) primary star and a low-mass star companion. The first of these substellar third bodies, a brown dwarf (BD), was announced roughly a decade ago around the post-common-envelope binary (PCEB) V471\,Tau \citep{guinan+ribas01-1}. At present, detection of substellar and mostly planetary circumbinary companions has been claimed for a dozen eclipsing, close-compact binaries containing WD or hot sdB primary stars \citep[see][ for the most convincing example]{beuermannetal10-1}. While the parameters derived for these potential circumbinary giant planets have to be considered with extreme caution, it is true that a large number of PCEBs show apparent period variations that might be explained by changes in the light travel time caused by a circumbinary giant planet. We present an observational census of eclipsing, close-compact binaries and eclipse timing measurements in the appendix.  

If confirmed, PCEBs as hosts of giant planets are particularly interesting because the binary host star passed through a special evolutionary phase: common-envelope (CE) evolution. As outlined by \citet{paczynski76-1} and described in much more detail by \citet{webbink84-1} CE evolution occurs once the initially more massive star (the primary) evolves. If it fills its Roche lobe during one of the giant phases, when it has a deep convective envelope, dynamically unstable mass transfer to the less massive component (the secondary) begins. The secondary is not able to adjust its structure on the mass-transfer time scale. The material lost by the primary therefore initially fills the Roche lobe of the secondary and then grows to form a noncorotating CE that surrounds the core of the giant and the secondary star. Owing to drag forces within the envelope, orbital energy is extracted from the binary and transferred to the envelope, which dramatically reduces the separation between the core of the primary and the secondary star in a spiraling-in process, until the envelope becomes unbound and is ejected from the binary. The remaining system is a PCEB consisting of the core of the primary (a compact object) and an MS companion, in a close but detached orbit. The short duration of the CE phase ($\lappr\,10^3$\,yr) means that the  mass of the secondary star is assumed to remain constant \citep{hjellming+taam91-1}, and the mass of the compact object should be equal to the mass of the core of the giant at the onset of mass transfer \citep[for more details of CE evolution see, e.g.,][]{iben+livio93-1,webbink08-1,zorotovicetal10-1}.

The special evolution of the host binary stars implies that the claimed circumbinary planets must have either survived the dramatic evolution of the host binary star or must have formed as a consequence of this evolution. Which of these two spectacular scenarios might have taken place in PCEBs with candidate circumbinary planets is currently a completely open question \citep[see, e.g., the discussion in][]{beuermannetal10-1}. Answering it based on hydrodynamic simulations of CE evolution, such as those presented by \citet{ricker+taam12-1}, is currently impossible, because CE evolution is still relatively uncertain even without an embedded planet.  

In this paper, we present the results of binary population synthesis simulations and reconstruct the evolutionary history of observed eclipsing PCEBs to characterize their MS+MS binary progenitors. Reviewing what is known about circumbinary planets and circumbinary planet formation around such MS+MS binaries, we conclude that the apparent period variations observed in PCEBs are very unlikely to be caused by giant planets that survived CE evolution, and we discuss alternative explanations. We start with a review of the observed sample of eclipsing PCEBs and the detected apparent period variations.  

\section{Giant planets around compact binaries?}\label{sec:census}

In the past few years, it has been suggested that there are giant planets around several eclipsing, close-compact binaries to explain observed variations of eclipse timings. These discoveries have become possible thanks to both the increase in the number of known eclipsing systems in recent years and the improvement in the accuracy of the observed light curves and derived eclipse timings. 

\addtocounter{table}{1}

Almost a decade ago, \citet{schreiber+gaensicke03-1} analyzed 30 detached PCEBs when only 11 eclipsing systems, including the prototype V471\,Tau, were known. The SDSS and intensive follow-up observations of WD+MS binaries, as well as the Catalina Realtime Transient Survey, significantly increased the number of known eclipsing, detached PCEBs. Table\,\ref{tab:sys} lists the orbital parameters of 56 known eclipsing, detached PCEBs and separates systems with WD and sdB primary stars. We excluded eclipsing PCEBs that are in the center of a planetary nebula, because no accurate eclipse timings are available for those systems. A brief review of the observational history of the 56 eclipsing PCEBs can be found in the appendix. 

Currently, 13 eclipsing, detached PCEBs consisting of an sdB (or sdOB) primary and a MS or BD companion are known. For an amazing fraction of $\sim\,38$ per cent, apparent period changes that might be caused by a third substellar body have been measured (systems in bold in Table\,\ref{tab:sys}). Even more dramatic, as shown in the appendix, almost all systems that have been intensively followed up for more than $\sim\,5$ years show period variations that might indicate the presence of a third body, with AA\,Dor the only exception \citep{kilkenny11-1}.

The sample of known eclipsing, detached PCEBs with a WD primary contains 43 systems. For four of them, third-body detections using eclipse timing measurements have been claimed. Very few eclipse timings have been published for almost all other systems, mostly because the systems have been discovered fairly recently. In addition, several early eclipse timings are not reliable. This means that, as in the case of sdB+MS eclipsing binaries, all systems with accurate eclipse timings (error $\lappr\,10$ sec) covering $\gappr\,5$ years show apparent period changes that may indicate there is a third circumbinary object.   
 
In the past few years, apparent period variations associated to the presence of a third body have also been observed for three cataclysmic variables (CVs): UZ\,For, HU\, Aqr, and DP\,Leo. CVs are PCEBs that evolved into a semi-detached configuration where a Roche lobe-filling MS star (or a BD) transfers mass on the WD. Eclipsing CVs were discovered more than a century ago \citep[see, e.g.,][]{pogson1857-1}, and up to now there are almost 200 eclipsing systems among the $\sim\,1000$ CVs listed in the catalog of \citet[][V7.16]{ritter+kolb03-1}. A complete discussion of all these systems is beyond the scope of this paper, but for completeness, we list the parameters of the three CVs with claimed planet detections in Table\,\ref{tab:sys} and briefly review the observational history of these three systems in the appendix.  

\addtocounter{table}{1}

The orbital parameters obtained with the best fit available for the claimed planets are listed in Table\,\ref{tab:plan}. It should be noticed that, at present, not a single set of orbital parameters has been confirmed by new measurements. In addition, some of the claimed planetary systems seem to be unstable \citep[see, e.g.,][]{horneretal11-1,hinseetal12-1}.  

Still, the large fraction of PCEBs that show apparent period variations is intriguing. Nine out of ten detached PCEBs, which corresponds to a fraction of $90\pm9$ per cent, with accurate eclipse timing measurements covering $\sim\,5$ years show clear variations that might indicate the presence of a third body. If these potential planets have formed prior to CE evolution, a similarly large fraction of the MS+MS progenitor binaries of PCEBs must also host giant planets. To evaluate whether this might indeed be the case, we need to derive the orbital and stellar parameters of the PCEB progenitors. 

\section{Binary population simulation}

To characterize the progenitors of PCEBs we performed binary population studies of PCEBs containing a WD or an sdB primary. We complement our simulations by reconstructing the evolutionary history of the observed sample using the code described in \citet{zorotovicetal11-1}.   

Binary population studies of WD+MS PCEBs have been previously performed by several authors \citep[see, e.g.,][]{dekool+ritter93-1,willems+kolb04-1,politano+weiler06-1,politano+weiler07-1,davisetal10-1}, and recently, \citet{clausenetal12-1} have simulated the population of sdB+MS binaries. We here present the first population simulation of PCEBs including both systems with WD and sdB primary stars. In  addition, it is the first time that the present-day population of PCEBs is simulated with a low value of the CE efficiency, including a fraction of the recombination energy, based on the results of \citet{zorotovicetal10-1}. A full parameter study will be presented elsewhere.

\subsection{Initial conditions and assumptions}

We use a Monte Carlo code to generate an initial population of $10^8$ MS+MS binaries. The mass of the primary star is distributed according to the initial mass function (IMF) of \citet{kroupaetal93-1}; i.e., the number of primaries with masses in the range $dM_\mathrm{1}$ is given by $dN \propto f(M_\mathrm{1})dM_\mathrm{1}$ where $f(M_\mathrm{1})$ is given by
\begin{equation}
f(M_\mathrm{1}) = \left\{\begin{array}{l l}
  0 & \quad \mbox{$M_\mathrm{1}/\Msun<0.1,$}\\
  0.29056M_\mathrm{1}^{-1.3} & \quad \mbox{$0.1\leq{M_\mathrm{1}/\Msun}<0.5,$} \\
  0.15571M_\mathrm{1}^{-2.2} & \quad \mbox{$0.5\leq{M_\mathrm{1}/\Msun}<1.0,$} \\
  0.15571M_\mathrm{1}^{-2.7} & \quad \mbox{$1.0\leq{M_\mathrm{1}/\Msun}.$} \\
  \end{array}
  \right.
\label{M1dist}
\label{eq:IMF}
\end{equation}
The mass of the secondary is assumed to be distributed according to a flat initial-mass-ratio distribution, i.e. $n(q)$ = constant, where  $q = M_\mathrm{2}/M_\mathrm{1}$. The initial orbital separation $a_\mathrm{i}$ follows the distribution
\begin{equation}
h(a_\mathrm{i}) = \left\{\begin{array}{l l}
  0 & \quad \mbox{$a_\mathrm{i}/\Rsun<3$ or $a_\mathrm{i}/\Rsun>10^{6},$}\\
  0.078636a_\mathrm{i}^{-1} & \quad \mbox{$3\leq a_\mathrm{i}/\Rsun \leq{10^6},$}\\ 
  \end{array}
  \right.
\label{adist}
\end{equation}
\citep{davisetal08-1}.
We also assign a ``born time'' ($t_{\mathrm{born}}$) to all the systems, corresponding to the age the galaxy had when the system was born, and assume that each $t_{\mathrm{born}}$  between $0$ and the age of the galaxy ($t_{\mathrm{gal}} \sim\,13.5$\,Gyr) is equally likely, which corresponds to assuming a constant star formation rate. 

Once the initial population has been generated, we use the latest version of the binary-star evolution (BSE) code from \citet{hurleyetal02-1} to evolve the systems. We have slightly updated the BSE code: the critical mass ratio for dynamically stable mass transfer when the primary is in the Hertzsprung gap was changed from 4.0 to 3.2 \citep{hanetal03-1}, and the rate of angular momentum loss due to magnetic braking was multiplied by the normalization factor provided by \citet{davisetal08-1}. Finally, we corrected a small mistake found by Robert Izzard (private communication)\footnote{Eq.\,~32 from \citet{hurleyetal02-1}, related to tidal effects in close binaries, was misspelled in the code. The correct equation is $f_{conv}=min \left[ 1,\left( \frac{P_{tid}}{2\tau_{conv}} \right)^2 \right]$, while in the code it was written  $f_{conv}=min \left[ 1,\left( \frac{(P_{tid})^2}{2\tau_{conv}} \right) \right]$. }

The systems are evolved for $t_{\mathrm{evol}} = t_{\mathrm{gal}} - t_{\mathrm{born}}$, to obtain the {\em{current}} orbital and stellar parameters. We assume a CE efficiency of 0.25 based on the results of \citet{zorotovicetal10-1} and the same fraction of recombination energy is included to compute the binding energy of the envelope. 

\subsection{Results}\label{sec:res}

After evolving the systems, we obtain more than $4 \times 10^5$ detached PCEBs consisting of a WD and a MS star or BD secondary. The mass and core composition of the WD clearly depend on the evolutionary stage of the primary star when the CE phase occurs.  

If the primary fills its Roche lobe during the first giant branch (FGB), the core will be mainly composed of helium. After the envelope is expelled, the core will not be massive enough to ignite helium and will evolve into a low-mass ($<0.5\Msun$), helium-core WD (He WD).  

If the CE phase occurs close to the tip of the FGB, the core of the primary may be massive enough to start helium burning following envelope ejection. The minimum core mass for which this occurs depends on the initial mass of the primary \citep[see, e.g.,][]{hanetal02-1}. When helium ignites after envelope ejection, the star becomes a hot sdB star, i.e. a helium-core-burning star with a very thin ($<0.02\Msun$) hydrogen envelope and a typical mass of $\sim\,0.45-0.5\Msun$. It remains in this stage for $\sim\,10^8$\,yr until helium is exhausted in the core. The low mass of the envelope means the star cannot ascend the asymptotic giant branch (AGB) and instead becomes a WD \citep[for a recent review of sdB stars see][]{heber09-1}. 

Finally if the primary fills the Roche lobe on the AGB the core of the primary will be more massive ($>0.5\Msun$), and it will become a carbon/oxygen-core WD (C/O WD) after the ejection of the envelope. Very massive primaries can also produce oxygen/neon-core WDs but for the sake of simplicity we here call all post-AGB WDs C/O WDs. 

The BSE code does not distinguish between He WDs and sdB stars; i.e., all the systems that enter the CE phase when the primary is on the FGB are assumed to evolve into PCEBs with He WD primaries. However, based on their current and initial masses, we can identify which of these systems are actually SdB stars: those with a mass higher than the minimum core mass ($M_\mathrm{c}^\mathrm{min}$) needed to ignite helium after the envelope is lost. To compute $M_\mathrm{c}^\mathrm{min}$ we here use the results from \citet[][ their Table\,~1]{hanetal02-1}, assuming a Reimers' wind mass-loss law (with $\eta$ = 1/4) and solar metallicity\footnote{\citet{hanetal02-1} only computed $M_\mathrm{c}^\mathrm{min}$ for two fixed values of metallicity (0.02 and 0.004). However, it can be seen from their Fig.\,~1 that the difference is not dramatic for stars with initial masses below $\sim\,1.8$ \Msun, which is the case for almost all the systems we obtain (see our Fig\,\ref{fig:miai})}.

We divide the PCEB systems into four types, depending on the type of primary  star they contain: \\
i) He WD primary, i.e., systems where the primary filled the Roche lobe on the FGB and $\Mwd < M_\mathrm{c}^\mathrm{min}$;\\
ii) sdB primary, i.e., systems where the primary filled the Roche lobe close to the tip of the FGB with $\Mwd \geq M_\mathrm{c}^\mathrm{min}$, which happened less than $10^8$\,yr ago;\\
iii) post-sdB primary, i.e., like the latter but if the CE phase occurred more than $10^8$\,yr ago, which means that the primary has been a helium burning sdB but has already evolved into a WD; \\
iv) C/O WD primary, i.e., systems that entered the CE phase when the primary was on the AGB. \\
Owing to the short duration of the sdB phase, we expect to find very few systems with primaries in this stage. 

It should be noticed that sdB stars can also be formed in binaries through stable mass transfer. However, their orbital periods are expected to be typically about two orders of magnitude larger than those of PCEBs \citep[e.g.,][]{podsiadlowskietal08-1}, so we do not take into account this formation scenario here.

\subsubsection{The predicted current PCEB population}

\begin{figure}
\centering
\includegraphics[angle=270,width=0.49\textwidth]{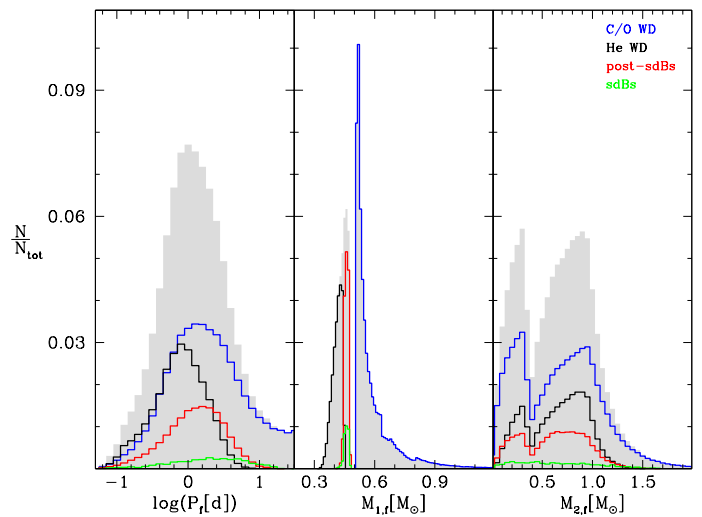}
\caption{Current orbital period (\textit{left}), primary mass (\textit{middle}), and secondary mass (\textit{right}) distribution of the simulated PCEB population, separated according to the type of primary star they contain. The gray shaded histograms represent the entire population. }
\label{fig:fin}
\end{figure}

Table\,\ref{tab:numb} summarizes the number of systems that are found in each of the four groups just described, as well as their average orbital parameters. The subscript ``f" corresponds to the final (current) value. As expected, very few ($\sim\,2\%$) systems with sdB primaries are still in this stage. The sample is dominated by systems with C/O WD primaries. The relative number of systems with each type of primary depends on the value assumed for the CE efficiency, but the shape of the distributions are not significantly affected \citep[see, e.g.,][]{willems+kolb04-1, politano+weiler07-1}. The resulting orbital period, primary mass, and secondary mass distributions of the predicted current PCEB population are shown in Fig.\,\ref{fig:fin}. The complete population of detached PCEBs is represented by the gray shaded histograms. Distributions separating different types of primaries stars are shown as color histograms. Since we have obtained a very low percentage of systems that still contain an sdB primary, we multiplied the number of systems in each bin by ten (in green). 

Systems in which the primary fills the Roche lobe at a more advanced evolutionary stage, which usually corresponds to a longer initial period, may
also end up with a longer final period. This is clearly reflected in the average final periods listed in Table\,\ref{tab:numb}, where the larger orbital periods are for systems containing C/O WDs, and is in perfect agreement with observations of WD+MS PCEBs \citep{zorotovicetal11-2}. However, it should be noticed that the the mean value ($\overline{P}_f$) of the period distribution for PCEBs containing C/O WDs is strongly affected by a huge tail towards very long orbital periods. This tail corresponds to systems that fill the Roche lobe very late on the AGB and whose orbital separation decreases very slightly during CE evolution owing to the large amount of recombination energy available within the envelope. This is also reflected in the high value of the standard deviation for the mean value. Therefore, we also give in Table\,\ref{tab:numb} the value of the median ($\widetilde{P}_f$), which is a better representation of the peak of the distribution and which still agrees with the results of \citep{zorotovicetal11-2}.

As expected, systems containing C/O WDs have $M_\mathrm{1,f}>0.5\Msun$, while progenitors that filled their Roche lobe on the FGB have $M_\mathrm{1,f}<0.5\Msun$ separated by a small gap. The lower boundary of this gap is given by the maximum core mass that a giant star can have at the end of the FGB ($\sim\,0.48$\Msun). The upper boundary appears because the radius of the star at the beginning of the AGB is smaller than at the end of the FGB. The minimum WD mass for a star that filled its Roche lobe during the AGB is given by the core mass at which the radius first exceeds the maximum FGB radius. This happens for core masses $\sim\,0.51$\Msun\footnote{These limits slightly depend on the metallicity.}. The mass distributions of He WDs and sdB (or post-sdB) stars overlap slightly, because the minimum core mass to ignite helium depends on the initial mass, but He WDs are in general slightly less massive than sdBs. 

The distribution of companion masses is similar for all types of PCEBs, which is also consistent with the observations. There are more and more systems with increasing secondary masses, with a steep decline at $M_2\sim\,0.35$\Msun, i.e. at the boundary for fully convective secondaries. According to the disrupted magnetic braking theory, systems with more massive secondaries suffer from angular momentum loss due to magnetic braking, in addition to gravitational radiation, and hence spend less time in the detached phase. This effect has been predicted theoretically by \citet{politano+weiler07-1} and observationally confirmed by \citet{schreiberetal10-1}. As expected, the orbital parameters for systems with sdB and post-sdB primaries are similar, with systems with a post-sdB primary having slightly shorter orbital periods since they have had more time to evolve towards shorter periods after the CE phase. 

\begin{table*}
\caption{\label{tab:numb} Current orbital properties of the simulated PCEB population separated by the type of primary star. The last column gives the median value for $P_f$, while all the others correspond to the means (see text for details). }
\begin{center}
\begin{tabular}{lccccccc}
\hline\hline
\noalign{\smallskip}
Primary & N            & Percent & $\overline{M}_{1,f}$ & $\overline{M}_2$ & $\overline{q}_f$ = $\overline{\left(\frac{M_2}{M_{1,f}}\right)}$ & $\overline{P}_f$ & $\widetilde{P}_f$ \\
        & $\times10^4$ & \%       & [\Msun]             & [\Msun]          &                                                     & [d]                & [d]\\
\hline
He WD    & 11.79 & 26.06 & 0.41 $\pm$ 0.02 & 0.65 $\pm$ 0.29 & 1.57 $\pm$ 0.71 & 1.03 $\pm$ 0.86 & 0.78\\
sdB      &  0.14 &  0.31 & 0.46 $\pm$ 0.01 & 0.65 $\pm$ 0.39 & 1.42 $\pm$ 0.86 & 3.23 $\pm$ 3.42 & 2.01\\
post-sdB &  6.87 & 15.18 & 0.46 $\pm$ 0.01 & 0.61 $\pm$ 0.31 & 1.34 $\pm$ 0.69 & 1.81 $\pm$ 1.69 & 1.32\\
C/O WD   & 26.45 & 58.45 & 0.59 $\pm$ 0.12 & 0.67 $\pm$ 0.42 & 1.15 $\pm$ 0.69 & 50.75 $\pm$ 154.94 & 2.54\\
\hline
\noalign{\smallskip}
\end{tabular}
\end{center}
\end{table*}

\subsubsection{Relating the initial separation and final system parameters} 

\begin{table*}
\caption{\label{tab:numbin} Initial primary mass, mass ratio, and orbital separation of the simulated PCEBs with the different primary types previously defined.}
\begin{center}
\begin{tabular}{lcccc}
\hline\hline
\noalign{\smallskip}
Primary & $\overline{M}_{1,i}$ & $\overline{q}_i$ = $\overline{\left(\frac{M_2}{M_{1,i}}\right)}$ & $\overline{a}_i$ & $\overline{a}_i$ \\
        & [\Msun] & & [\Rsun] & [AU]\\
\hline
He WD    & 1.20 $\pm$ 0.15 & 0.53 $\pm$ 0.23 & 238 $\pm$ 53 & 1.11 $\pm$ 0.25 \\
sdB      & 1.32 $\pm$ 0.23 & 0.48 $\pm$ 0.26 & 361 $\pm$ 51 & 1.68 $\pm$ 0.24 \\
post-sdB & 1.39 $\pm$ 0.19 & 0.45 $\pm$ 0.22 & 358 $\pm$ 45 & 1.67 $\pm$ 0.21 \\
C/O WD   & 2.08 $\pm$ 0.93 & 0.34 $\pm$ 0.21 & 699 $\pm$ 356& 3.25 $\pm$ 1.65 \\
\hline
\noalign{\smallskip}
\end{tabular}
\end{center}
\end{table*}

\begin{figure}
\centering
\includegraphics[angle=270,width=0.49\textwidth]{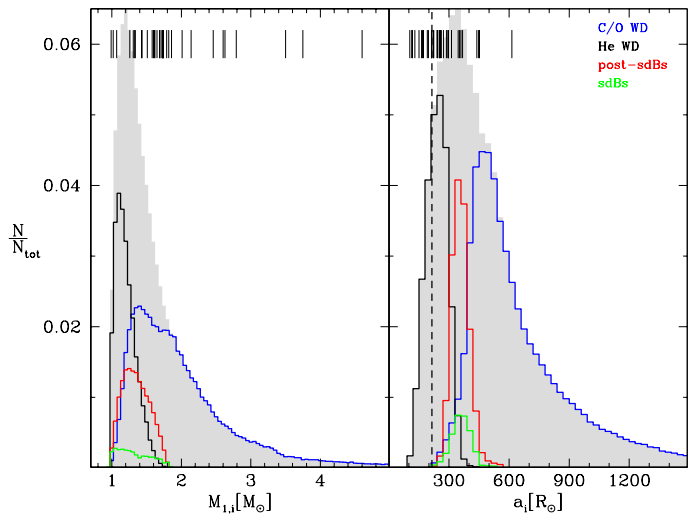}
\caption{Initial mass distribution of the primary (\textit{left}) and initial separation (\textit{right}) for the different types of PCEBs obtained with our simulation. The short vertical lines in the upper part of each panel correspond to the reconstructed values for the systems in Table\,\ref{tab:sys}. The dashed vertical line corresponds to $1$\,AU.}
\label{fig:miai}
\end{figure}

In the context of detecting circumbinary planets around PCEBs, the initial parameters of the systems, and their relation with the current parameters is crucial. 

One of the basic assumptions for the CE phase is that the mass of the secondary star remains constant, therefore the initial secondary masses are almost identical to the current secondary masses. The distributions of initial primary masses and initial separations are shown in Fig.\,\ref{fig:miai}, and the mean values are listed in Table\,\ref{tab:numbin}. Again, systems with sdB and post-sdB primaries cover the same range of initial parameters. As in Fig.\ref{fig:fin} the number of systems with sdB primaries was multiplied by ten in Fig.\,\ref{fig:miai}.  

All the primaries that fill their Roche lobe on the FGB, i.e. the progenitors of PCEBs with sdB and He-core WD primaries, descend from low-mass stars ($M_\mathrm{1,i}\,\lappr\,1.8\Msun$), while the progenitors of C/O WDs are generally more massive and cover a wider range of masses.  

Apparently, most PCEBs have formed from binaries with relatively small initial separations $a_i\,\lappr\,1000$\Rsun, i.e. $\lappr\,5$\,AU. The dashed vertical line in the right-hand panel of Fig.\,\ref{fig:miai} represents $1$\,AU for comparison. PCEBs containing He WD primaries descend from very close binaries with $a_i= 100-400$\Rsun, the progenitors of PCEBs with an sdB primary had initial separations of $200-500$\Rsun, and PCEBs with C/O WD primaries formed mostly from systems with initial separations $\gappr\,500\Rsun$. This is not surprising, because the radius of a giant star increases as the star evolves and the mass of the core grows, and therefore, larger separations are needed for the Roche lobe to be filled with a more massive core. 

\begin{figure*}
\centering
\includegraphics[angle=270,width=0.7\textwidth]{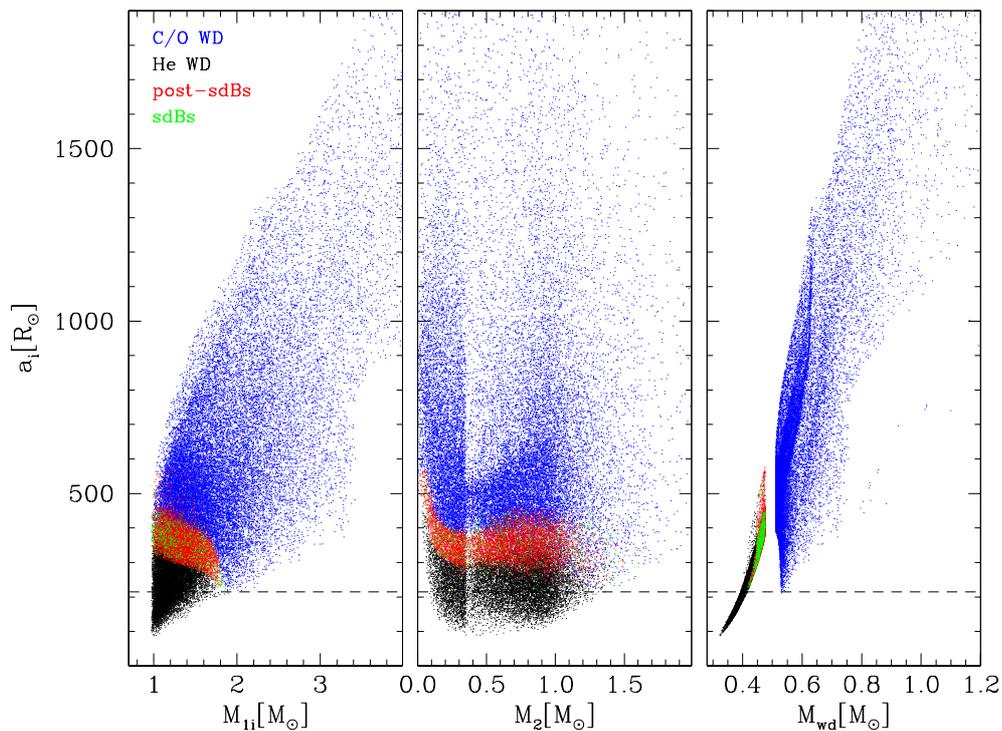}
\caption{Relation between the initial orbital separation and: initial primary mass (\textit{left}), secondary mass (\textit{middle}), and current mass of the primary (\textit{right}).}
\label{fig:am}
\end{figure*}

The relation between the initial binary separation and the initial and final stellar masses is shown in Fig.\,\ref{fig:am}. There seems to be a relation with the initial primary mass and, more evidently, with the final WD mass. This is a consequence of the radius of a giant star depending almost exclusively on the core mass, i.e. on the mass of the future WD. This implies that the initial separation should also be larger in order to allow the star to evolve to this stage without filling the Roche lobe. 

To compare the predictions of our simulations with the observed sample of eclipsing PCEBs, we use the reconstruction algorithm presented in \citet{zorotovicetal11-1} to obtain the initial orbital parameters for all the systems in Table\,\ref{tab:sys} with available periods and masses. The resulting values are included in Fig.\,\ref{fig:miai} in the upper part of each panel\footnote{We can only calculate the initial separations at the onset of the CE phase, which might differ from the actual initial separation due to, e.g., mass loss and tidal effects that may affect the orbital separation prior to CE evolution.}. The sample of observed eclipsing PCEBs is naturally biased towards short orbital periods, hence also slightly towards small initial binary separations of just a few AU, but otherwise the properties of the simulated and reconstructed PCEBs are similar. 

In summary, we found that PCEBs in general and the observed population of eclipsing PCEBs in particular descend from binaries with primary masses of $M_1\sim\,1-3\Msun$, secondary masses from $0.1-1.0\Msun$ and initial separations of $a_i\,\lappr\,5$\,AU. 

\section{Planet formation and planets around MS+MS binaries}

In the previous sections we have derived the properties of the MS+MS binary progenitors of PCEBs. In what follows we review what is known about planet formation and circumbinary planets around such MS+MS binaries. We start with a brief review of planet formation theories. 

\subsection{Giant planet formation}\label{sec:models} 

Giant planet formation theory has been developed early for the solar system and the pioneering works of the two still today competing models, i.e. the core-accretion model \citep{safranov69-1,goldreich+ward73-1} and the disk-instability model \citep{cameron78-1}, were published well before the first exoplanets were identified. 

According to the disk-instability scenario, massive protoplanetary disks may fragment into dense cores that can contract further and form giant planets. Gravitational instabilities occur if Toomre's Q parameter \citep{toomre64-1} approaches unity and if the local cooling time is shorter than or equal to the orbital period at a given radius. Current models almost concordantly (to the best of our knowledge, the only exception is \citealt{boss12-1}) show that such conditions might be present in massive young ($\lappr\,1$\,Myr) disks and only at distances from the central star $\gappr\,20-100\,$AU \citep[e.g.,][]{boley09-1,caietal10-1}. However, migration of the planet and scattering can redistribute the planetary orbits, and one can therefore not exclude relatively close planets also being formed due to disk instabilities. 
 
According to the alternative scenario, the core-accretion model, larger protoplanetary cores are built by the accretion of planetesimals until the core is massive enough to accrete gas from the surrounding protoplanetary disk. The accretion of planetesimals is relatively slow, and the formation of giant planets according to the core-accretion models takes around four million years \citep{alibertetal05-1}. The strengths of the core-accretion model are that it naturally explains the planet-metallicity relation \citep{fischer+valenti05-1}, that gas giants formed by core accretion are enriched due to the accretion of planetesimals as was been found to be the case in our solar system \citep{young03-1}, and that rocky planets can form due to basically the same mechanism. Furthermore, population models based on the core-accretion scenario predict mass and semi-major axis distributions in agreement with observations \citep{ida+lin04-1,alibertetal05-1}. Only very few massive planets around low-metallicity stars might be difficult to be explained within the core-accretion scenario \citep{mordasinietal12-1}. The core-accretion model is  currently the favored formation model for most giant planets although giant planet formation due to disk instabilities cannot be excluded.

\subsection{Multiplicity and exoplanets}\label{sec-pft} 

Already almost two decades ago, \citet{ghezetal93-1} were the first to find a difference between the binary separation distributions of classical accreting T\,Tauri stars (CTTS) and disk-less weak-lined T\,Tauri stars (WTTS), in the sense that the fraction of WTTS to CTTS is increased for short projected separations $\lappr\,50$\,AU. This finding has meanwhile been impressively confirmed by several studies using the unprecedented database provided by the Spitzer survey \citep[e.g.,][]{ciezaetal09-1}, and similar results are found for the transition disk \citep{ciezaetal12-2} and debris disk systems \citep{rodriguez+zuckerman12-1}. The lack of binary stars with projected binary separations $\sim\,1-100$AU among samples of young stellar objects still hosting a disk may indicate a short lifetime of circumbinary disks around these close binaries.

Indeed, very recently, \citet{krausetal11-1,krausetal12-1} have presented results of a high-resolution imaging study of the Taurus Auriga star-forming region of mostly solar type stars ($M\sim\,0.25-2.5\Msun$) and compared the fraction of pre-MS stars hosting disks at different ages. They clearly show that the paucity of binaries among CTTS, transition disks, and debris disks is explained by the short lifetime of most circumbinary protoplanetary disks, i.e. $\lappr\,1$\,Myr. As this is short compared to the time scale of giant planet formation according to the core-accretion model, only a small fraction of these binaries should host circumbinary planets if core accretion is indeed the main channel for forming giant planets.  

In addition to the short lifetimes of circumbinary disks around solar-type binary stars with projected separations $\gappr\,1$\,AU, the search of planets around MS+MS binary stars within the field of NASA's Kepler mission revealed a low frequency of giant planets around even closer solar-type MS+MS binaries. The first planet around a pair of MS stars, Kepler-16, was found by \citet{doyleetal11-1}. Later, \citet{welshetal12-1} found two more planet transits-like events around the MS+MS binaries Kepler-34 and Kepler-35, in a survey of 750 eclipsing binaries with periods ranging from 0.9 to 276 days, from Kepler's survey of MS+MS binaries. These three systems have binary periods longer than 20 days, while most of the systems observed by Kepler ($>80\%$) have shorter orbital periods. Considering that it is more likely to detect planets around short-period systems, this suggests a much lower rate of gas giant planets around the shorter period binaries. Based on geometric arguments, \citet{welshetal12-1} infer a lower limit of $\sim\,1\%$ of circumbinary giant planets with periods around 100--200 days in close binaries (P = 0.9-50 days). 

In summary, the low frequency of circumbinary planets complements the apparently short lifetimes of circumbinary disks for projected separations of $\gappr\,1$AU. We conclude that evidence for small numbers of circumbinary giant planets around close solar-type binary stars is growing. Such binary stars are the progenitors of PCEBs, so that the high frequency of apparent period variations observed in eclipsing PCEBs (see Sect.\,\ref{sec:census}) is unlikely to be caused by giant planets that survived CE evolution. In fact, the only remaining possibility to explain the period variations observed in nearly all PCEBs due to such first-generation planets requires that nearly all first-generation circumbinary disks form giant planets at large separations that are not yet detectable with Kepler. Furthermore, these planets must form  due to the fast disk-instability mechanism that is not in conflict with the short disk lifetimes mentioned above. While current observations cannot entirely exclude this possibility, it appears unlikely given that a flat extrapolation of the observed planet separation distribution to separations of $3-10$\,AU around single stars predicts that only $\sim\,12$ per cent may host giant planets \citep{marcyetal05-1}.

\section{Discussion}

In the previous sections we obtained the following results

\begin{itemize}

\item From the 56 eclipsing, detached PCEBs with MS or BD secondaries, accurate eclipse time measurements covering more than around five years have been obtained for ten systems. All of these but AA\,Dor show apparent period changes that might indicate the presence of circumbinary planets.

\item The progenitors of PCEBs are MS+MS binaries with initial separations of $\sim\,0.5-5$\,AU, similar to those tested by \citet{krausetal12-1} and \citet{welshetal12-1}. 

\item The reduced lifetime of circumbinary protoplanetary disks does not allow the frequent formation of circumbinary planets around close MS+MS binaries \citep{krausetal12-1} within the core-accretion model. This has been shown for projected separations $\sim\,1-40$\,AU and is likely to also be true for closer binaries \citep{welshetal12-1}.    

\end{itemize} 

These results suggest that the apparent period changes in PCEBs are not likely to be caused by circumbinary giant planets that formed in the circumbinary protoplanetary disk around the initial MS+MS binary. Instead, if the variations are indeed caused by circumbinary planets, the planets may have formed following CE evolution in a second-generation disk. The idea of second-generation planets goes back to the first exoplanet discovered around the pulsar PSR\,B1257+12. According to \citet{tavani+brookshaw92-1}, a circumbinary protoplanetary disk may have formed around the pulsar from evaporated material that did not leave the system. More recent models start from supernova fallback disks and can reproduce the observed eccentricities and masses of the planetary system around PSR\,B1257+12 \citep{hansenetal09-1}. It is therefore reasonable to suggest that also the claimed planets around PCEBs have a second-generation origin. In such a scenario, a protoplanetary disk must have formed following the CE ejection. This idea is supported by recent simulations of CE evolution indicating that envelope ejection can be incomplete, i.e. up to $\sim\,10$ per cent of the ejected material may remain bound to the PCEB and is likely to form a circumbinary disk \citep{kashi+soker11-1,ricker+taam12-1}. If the apparent period changes in PCEBs are indeed caused by planets formed in such a second-generation protoplanetary disk, the planet formation process must proceed on a short time scale since the cooling age of the most convincing PCEB with planets, NN Ser, is just a million years. It must also be very efficient because nearly all PCEBs with accurate eclipse timings show apparent period variations. Below we discuss several options of second-generation planet formation and add a note of caution concerning the third-body interpretation.    

\subsection{Second-generation disk instabilities}

The disk-instability model offers planet formation on short time scales that could solve the disagreement with the circumbinary planets around very young PCEBs such as NN\,Ser. However, in general, the disk-instability model still has to overcome serious problems, as recently discussed by \citet{zhuetal12-1}. The orbital separations of the claimed planet detections around PCEBs disagree with the predictions of the disk instability model. Almost all the claimed planets are in orbits with $a\,\lappr\,10$ AU (see Table\,\ref{tab:plan}), while the conditions for forming planets according to the disk-instability scenario occur at distances $\gappr\,20$ \,AU from the central star. Significant planet migration and/or scattering would thus be necessary. Furthermore, given the high fraction of PCEBs showing period variations would require that nearly all post-CE circumbinary disk are unstable and form giant planets. Although this scenario cannot be excluded, it is not clear why second-generation disks should be almost always unstable, fragment, and form giant planets, while disks around young single stars and MS+MS binaries seem to form giant planets in only about ten per cent of all cases. Therefore, second-generation disk instabilities do not appear to be the most promising solution.

\subsection{Second-generation rapid core accretion}

Planet formation according to the core-accretion model is more likely for higher metallicities \citep[e.g.][]{fischer+valenti05-1}, and therefore the production of massive elements during the AGB evolution might favor the formation of second-generation giant planets. Indeed, dusty disks surrounding binary post-AGB stars are a well known phenomenon. \citet{watersetal98-1} were the first to claim that such disks are similar to protoplanetary disks and may lead to the formation of planets. More recently, evidence has grown that dust-processing in these disks and in protoplanetary disks around young stars is probably very similar \citep{gielen08-1}. Moreover, a debris second-generation disk around an isolated neutron star has been detected \citep{wangetal06-1}. Furthermore, the second-generation giant planet formation process might be fast if the envelope that forms the second-generation disk is rich of heavy elements, especially C and O, that trigger enhanced dust production and planet formation \citep{petigura+geoffrey11-1}. Significantly enhanced dust-to-gas mass ratios may be able to explain both the high frequency of circumbinary planets and the short time scale of the second-generation planet formation required to explain the youth of the WD in NN\,Ser ($\sim\,1$\,Myr). Therefore, the second-generation, rapid core-accretion scenario seems to be a reasonable option to explain the observed apparent period variations.  

Even better, based on the above, the second-generation disk model makes predictions that can be tested observationally. Enrichment of the envelope with heavy elements is well established as occurring at the end of the AGB but should be completely absent on the early FGB. As a result, second-generation disks around PCEBs with high-mass C/O WD primaries should have much higher metallicities than around systems with low-mass He WD primaries and should therefore be found to more frequently host giant planets\footnote{However, this prediction is only based on metallicity, which might not be the only difference in the formation of C/O and He-core WDs that could affect potential planet formation in a second-generation circumbinary disk. Other factors, such as the expelled mass that is higher for AGB progenitors, may complicate the issue and dilute the prediction.}.

A borderline case is PCEBs with sdB primary stars. Some authors claim that dust production may also occur very close to the tip of the FGB \citep{boyeretal10-1}, precisely where the progenitors of sdBs fill their Roche lobes. The evolution towards sdB stars is far from being completely understood. Late He-flashes may well lead to mixing in systems that experienced significant mass loss, but details are missing. We know neither the details of CE evolution nor pulsations or late He-flashes that may occur at the tip of the FGB. All we know for certain is that sdB stars are burning He, which produces C/O, and one may speculate that this leads to enhanced metallicities in a potential second-generation disk around PCEBs with sdB primary stars.    

The current data is inconclusive with respect to possible relations between the nature of the compact object and occurrence of circumbinary planets. There are nine PCEBs and three CVs with period variations attributed to circumbinary planets (see Table\,\ref{tab:sys}). Among the PCEBs, five of them contain an sdB primary, three a C/O WD, and only one a WD with $\Mwd\,<0.5$. For the last, RR\,Cae, apparent period variations have been only recently detected, and still need to be confirmed. Also, the mass of the WD is high enough to be either a He WD or a post-sdB star. In the case of CVs, all of them contain C/O WDs, which is not surprising given the observed mass distribution of WDs in CVs \citep{zorotovicetal11-1}. These fractions agree with the second-generation disk scenario but are still far from providing clear constraints. Intense monitoring of a large sample of eclipsing PCEBs is required to test the predictions of second-generation planet formation. 

\subsection{A hybrid first- and second-generation scenario}

Neptune-mass planets seem to be more frequent than giant Jupiter-like planets \citep{boruckietal11-1}, at least around single stars. If this can be further confirmed, especially for planets around close MS+MS binary stars, it might be that second-generation planet formation is efficient because the remnants of lower mass planets survived CE evolution and serve as seeds for more massive planets. As these planets already have a few Earth masses, the formation of a giant planet might be faster than for first-generation giant planets, and the short cooling age of the WD in some PCEBs is no longer a problem. The predictions of this scenario are a high fraction of low-mass planets around MS+MS binaries, which subsequently can grow due to the accretion of gas left by the CE phase. If this is the case, there should be no relation between the type of compact object in PCEBs and the frequency of circumbinary planets. While low-mass circumbinary planets around MS+MS binaries have indeed recently been detected \citep{oroszetal12-1,oroszetal12-2}, proper statistics of the frequency of low-mass circumbinary planets are not yet available.

\subsection{A note of caution: are we really detecting planets?}

As outlined in the introduction, pulsar timings led to detecting the first confirmed exoplanet orbiting the pulsar PSR\,B1257+12 \citep{wolszczan+frail92-1,wolszczan94-1}, so planets around compact binary stars do exist. However, one also has to keep in mind that the possible companion to PSR\,1829-10 \citep{bailesetal91-1} was retracted later \citep{lyne+bailes92-1}.   

The situation for PCEBs, and especially CVs, is similar. All we see are eclipse timing variations that can be reproduced well by assuming the existence of an orbiting circumbinary third body, but not a single set of eclipse times that confirms the parameters of a previously claimed planetary systems has been published so far. In contrast, some suggested planetary systems \citep[e.g.,][]{qianetal10-2,qianetal11-1} turned out to be dynamically unstable \citep{horneretal11-1,hinseetal12-1} or drastically disagreed with more recent high-precision eclipse timings \citep{parsonsetal10-2}. It therefore remains an open question whether we indeed identified a new, large, and exciting population of circumbinary extrasolar planets or if perhaps an alternative process, such as the frequently proposed Applegate's mechanism \citep{applegate92-1} or a so far unknown process acting in deeply convective secondary stars might be responsible for at least some of the observed timing variations. In this context, AA\,Dor, a PCEB that may contain a BD companion, is extremely interesting because it is so far the only PCEB with continuous high-precision eclipse time measurements that does not show any signs of apparent period variations \citep{kilkenny11-1}. One might therefore speculate that indeed the apparent period changes are somehow related to the existence of the convective secondary present in all PCEBs and CVs with claimed planet detections instead of being caused by circumbinary planets. Observations of eclipsing PCEBs with other types of secondary stars, i.e. with a WD or a BD companion (like e.g., SDSSJ0820+0008), may give new insight into this alternative possibility.

\section{Conclusion}

By combining binary population models with recent observational and theoretical results for the formation of circumbinary giant planets, we have shown that the apparent period variations seen in virtually all close-compact binaries with good coverage ($\gappr\,5$\,yr of mid-eclipse times with good accuracy) are unlikely to be explained by first-generation planets. The lifetimes of protoplanetary disks around MS+MS binaries are simply too short to form giant planets in most cases \citep{krausetal12-1}, and the observed fraction of planets around MS+MS stars further confirms this \citep{welshetal12-1}. The remaining options for explaining the observed period changes are either second-generation planet formation or alternative explanations that do not involve the existence of third and fourth bodies. We proposed observational experiments to test both hypothesis.   

First, if not caused by circumbinary planets but the active secondary, the period variations should not be detected in close WD binaries with a second WD component or with a BD secondary. Second, if caused by second-generation planets, a clear relation between dust production in the envelope of the compact object progenitor and planet frequency is expected; i.e., planets should be more frequent around compact binaries with C/O WDs than around those with a He-core WD primary. Continuous monitoring of all the eclipsing PCEBs listed in the appendix and also of all new systems that will be discovered will shed light on the origin of the observed apparent period changes in PCEBs.

\begin{acknowledgements}
MZ acknowledges support from comite mixto, Gemini/Conicyt (grant 32100026) and CONICYT/FONDECYT/POSTDOCTORADO/3130559. MRS thanks FONDECYT (project 1100782) and the Millennium Science Initiative, Chilean Ministry of Economy, Nucleus P10-022-F. We thank Steven G. Parsons for helpful discussions.  
\end{acknowledgements}

\longtab{1}{
\begin{longtable}{llcccl}
\caption{\label{tab:sys} Orbital parameters for the currently known eclipsing PCEBs. Systems with suspected planets are in bold. We distinguish between detached systems containing sdB primaries, detached systems with WD primaries, and CVs. For the case of CVs we only list those with suspected planets (see Sect.\,\ref{sec:census} for details).}\\
\hline\hline
\noalign{\smallskip}
System  & Alt. Name & $\Porb$ &  $M_1$    &    $M_2$   & References\\
        &           & [d]     &  [\Msun]  &   [\Msun]   &  \\
\noalign{\smallskip}
\hline
\endhead
\noalign{\smallskip}
\hline
\endfoot
\noalign{\smallskip}
\multicolumn{6}{l}{Detached sdB+MS/BD PCEBs}	\\
\hline
\noalign{\smallskip}
\textbf{HW\,Vir}        & \textbf{2M\,J1244-0840}  & \textbf{0.11671955} & \textbf{0.485 $\pm$0.013} & \textbf{0.142$\pm$0.004}& 1,2          \\
\textbf{HS\,0705+6700}  & \textbf{2M\,J0710+6655}  & \textbf{0.095646625} & \textbf{0.483}            & \textbf{0.134}          & 3, 4      \\
\textbf{HS\,2231+2441}  & \textbf{2M\,J2234+2456}  & \textbf{0.110588}   & \textbf{0.47:}            & \textbf{0.075:}         & 5         \\
\textbf{NSVS\,14256825} & \textbf{2M\,J2020+0437}  & \textbf{0.1103741}  & $\sim\,$\textbf{0.46}       & $\sim\,$\textbf{0.21}     & 6   	    \\
\textbf{NY\,Vir}        & \textbf{PG\,1336-018}    & \textbf{0.101015967}& \textbf{0.459$\pm$0.005}  & \textbf{0.122$\pm$0.001}& 7, 8, 9 \\
2M1938+4603             & NSVS\,05629361           & 0.1257653           & 0.48$\pm$0.03             &    0.12$\pm$0.01        & 10         \\
NSVS\,07826247          & CSS06833                 & 0.16177042          & 0.376$\pm$0.055           &    0.113$\pm$0.017      & 11         \\
BUL-SC16 335            & 2M\,J1809-2641           & 0.12505028          & 0.5:                      &    0.16:                & 12         \\
PG\,1621+4737           & 2M\,J1622+4730           & 0.075               & -                         &    -                    & 13         \\
SDSSJ0820+0008          & GSC\,0196.0617           & 0.097               & $\sim\,$0.25              &    0.045$\pm$0.03       & 14, 15     \\
			& 		           &                     & $\sim\,$0.47              &    0.068$\pm$0.03       & 15     \\
ASAS\,10232             & 2M\,J1023-3736           & 0.13927             & 0.461$\pm$0.051           &    0.157$\pm$0.017      & 16         \\
AA\,Dor                 & LB\,3459                 & 0.261539736         & 0.471$\pm$0.005           &$0.0788^{+0.0075}_{-0.0063}$& 17,18    \\
EC\,10246-2707          &                          & 0.118507993         & 0.45$\pm$0.17             &    0.12$\pm$0.05        & 19    \\
\noalign{\smallskip}
\hline
\noalign{\smallskip}
\multicolumn{6}{l}{Detached WD+MS PCEBs} 														    \\
\hline
\noalign{\smallskip}
\textbf{NN\,Ser}        & \textbf{2M\,J1552+1254}  & \textbf{0.13008014} & \textbf{0.535$\pm$0.012}  & \textbf{0.111$\pm$0.004}& 20, 21     \\
\textbf{V471\,Tau}      & \textbf{2M\,J0350+1714}  & \textbf{0.52118343} & \textbf{0.84$\pm$0.05}    & \textbf{0.93$\pm$0.07}  & 22, 23     \\
\textbf{QS\,Vir}        & \textbf{EC\,13471-1258}  & \textbf{0.1507575}  & \textbf{0.78$\pm$0.040}   & \textbf{0.430$\pm$0.040}& 24         \\
\textbf{RR\,Cae}        & \textbf{2M\,J0421-4839}  & \textbf{0.30370363} & \textbf{0.440$\pm$0.022}  & \textbf{0.183$\pm$0.013}& 25         \\
DE\,Cvn                 & RX\,J1326.9+4532         & 0.364139315         & $0.51^{+0.06}_{-0.02}$    &    0.41$\pm$0.06        & 26, 27     \\
GK\,Vir                 & SDSSJ1415+0117           & 0.344330833         & 0.564$\pm$0.014           &    0.116$\pm$0.003      & 28         \\
RX\,J2130.6+4710        & 2M\,J2130+4710           & 0.52103562          & 0.554$\pm$0.017           &    0.555$\pm$0.023      & 29         \\
SDSSJ0110+1326          & WD\,0107+131             & 0.332687            & 0.47$\pm$0.02             &    0.255-0.380          & 30         \\
SDSSJ0303+0054          &                          & 0.1344377           & 0.878-0.946               &    0.224-0.282          & 30         \\
SDSSJ0857+0342          & CSS03170                 & 0.06509654          & 0.51$\pm$0.05             &    0.09$\pm$0.01        & 31         \\
SDSSJ1210+3347          &                          & 0.12448976          & 0.415$\pm$0.010           &    0.158$\pm$0.006      & 32         \\
SDSSJ1212-0123          &                          & 0.3358711           & 0.439$\pm$0.002           &    0.273$\pm$0.002      & 27         \\
SDSSJ1435+3733          &                          & 0.125631            & 0.48-0.53                 &    0.190-0.246          & 30         \\
SDSSJ1548+4057          &                          & 0.1855177           & 0.614-0.678               &    0.146-0.201          & 30         \\
CSS06653                & SDSSJ1329+1230           & 0.08096625          & 0.350$\pm$0.081           &    -                    & 33, 34     \\
CSS07125                & SDSSJ1410-0202           & 0.363497            & 0.470$\pm$0.055           &    0.380$\pm$0.012      & 34, 35     \\
CSS080408               & SDSSJ1423+2409           & 0.3820040           & 0.410$\pm$0.024           &    0.255$\pm$0.040      & 34, 35     \\
CSS080502               & SDSSJ0908+0604           & 0.14943807          & 0.370$\pm$0.018           &    0.319$\pm$0.061      & 33, 34     \\
CSS09704                & SDSSJ2208-0115           & 0.1565057           & 0.37                      &    -                    & 35         \\
CSS09797                & SDSSJ1456+1611           & 0.229120            & 0.370$\pm$0.016           &    0.196$\pm$0.043      & 34, 35     \\
CSS21357                & SDSSJ1348+1834           & 0.2484              & 0.590$\pm$0.017           &    0.319$\pm$0.061      & 34, 35     \\
CSS21616                & SDSSJ1325+2338           & 0.1949589           & -                         &    -                    & 33         \\
CSS25601                & SDSSJ1244+1017           & 0.227856            & 0.400$\pm$0.026           &    0,319$\pm$0.061      & 34, 35     \\
CSS38094                & SDSSJ0939+3258           & 0.3309896           & 0.520$\pm$0.026           &    0.319$\pm$0.061      & 33, 34     \\
CSS40190                & SDSSJ0838+1914           & 0.13011232          & 0.390$\pm$0.035           &    0.255$\pm$0.040      & 33, 34     \\
CSS41631                & SDSSJ0957+2342           & 0.15087074          & 0.430$\pm$0.025           &    0.431$\pm$0.108      & 33, 34     \\
WD\,1333+005            & SDSSJ1336+0017           & 0.1219587           & -                         &    -                    & 33         \\
PTFEB11.441             & PTF1\,J004546.0+415030.0 & 0.3587              & 0.51$\pm$0.09             &    0.35$\pm$0.05        & 36         \\
PTFEB28.235             & PTF1\,J015256.6+384413.4 & 0.3861              & 0.65$\pm$0.11             &    0.35$\pm$0.05        & 36         \\
PTFEB28.852             & PTF1\,J015524.7+373153.8 & 0.4615              & 0.52$\pm$0.05             &    0.35$\pm$0.05        & 36         \\
KIC-10544976            & USNO-B1.0\,1377-0415424  & 0.35046872          & 0.61$\pm$0.04             &    0.39$\pm$0.03        & 37         \\
SDSS J0821+4559         &                          & 0.50909             & 0.66$\pm$0.05             &    0.431$\pm$0.108      & 34, 38     \\
SDSS J0927+3329         &                          & 2.30822             & 0.59$\pm$0.05             &    0.380$\pm$0.012      & 34, 38     \\
SDSS J0946+2030         &                          & 0.252861219         & 0.62$\pm$0.10             &    0.255$\pm$0.040      & 34, 38     \\
SDSS J0957+3001         &                          & 1.92612             & 0.42$\pm$0.05             &    0.380$\pm$0.012      & 34, 38     \\
SDSS J1021+1744         &                          & 0.14035907          & 0.50$\pm$0.05             &    0.319$\pm$0.061      & 34, 38     \\
SDSS J1028+0931         &                          & 0.23502576          & 0.42$\pm$0.04             &    0.380$\pm$0.012      & 34, 38     \\
SDSS J1057+1307         &                          & 0.1251621           & 0.34$\pm$0.07             &    0.255$\pm$0.040      & 34, 38     \\
SDSS J1223-0056         &                          & 0.09007             & 0.45$\pm$0.06             &    0.196$\pm$0.043      & 34, 38     \\
SDSS J1307+2156         &                          & 0.216322132         & -                         &    0.319$\pm$0.061      & 34, 38     \\
SDSS J1408+2950         &                          & 0.1917902           & 0.49$\pm$0.04             &    0.255$\pm$0.040      & 34, 38     \\
SDSS J1411+1028         &                          & 0.167509            & 0.36$\pm$0.04             &    0.380$\pm$0.012      & 34, 38     \\
SDSS J2235+1428         &                          & 0.14445648          & 0.45$\pm$0.06             &    0.319$\pm$0.061      & 34, 38     \\
\noalign{\smallskip}
\hline
\noalign{\smallskip}
\multicolumn{6}{l}{CVs}														    \\
\hline
\noalign{\smallskip}
\textbf{UZ\,For}        & \textbf{2M\,J0335-2544}  & \textbf{0.08786542} & $\sim\,$\textbf{0.71}       & $\sim\,$\textbf{0.14}     & 39, 40       \\
\textbf{HU\,Aqr}        & \textbf{2M\,J2107-0517}  & \textbf{0.08682041} & \textbf{0.80$\pm$0.04}    & \textbf{0.18$\pm$0.06}  & 41, 42       \\
\textbf{DP\,Leo}        & \textbf{RX\,J2107.9-0518}& \textbf{0.06236286} & \textbf{1.2:}             & \textbf{0.14:}          & 43, 44       \\
                        &                          &                     & \textbf{0.6:}             & \textbf{0.09:}          & 45        
\end{longtable} \textit{References}.
(1)~\citet{leeetal09-1},
(2)~\citet{beuermannetal12-2},
(3)~\citet{drechseletal01-1},
(4)~\citet{beuermannetal12-1},
(5)~\citet{ostensenetal07-1},
(6)~\citet{wilsetal07-1},
(7)~\citet{vuckovicetal07-1},
(8)~\citet{charpinetetal08-1},
(9)~\citet{qianetal11-2},
(10)~\citet{ostensenetal10-1},
(11)~\citet{foretal10-1},
(12)~\citet{polubeketal07-1},
(13)~\citet{geieretal10-1},
(14)~\citet{geieretal11-4},
(15)~\citet{geieretal11-1},
(16)~\citet{schaffenrothetal11-1},
(17)~\citet{kilkenny11-1},
(18)~\citet{klepp+rauch11-1},
(19)~\citet{barlowetal12-1},
(20)~\citet{parsonsetal10-1},
(21)~\citet{beuermannetal10-1},
(22)~\citet{obrienetal01-1},
(23)~\citet{kundra+hric11-1},
(24)~\citet{odonogueetal03-01},
(25)~\citet{maxtedetal07-2},
(26)~\citet{vandenbesselaaretal07-1},
(27)~\citet{parsonsetal10-2},
(28)~\citet{parsonsetal12-1},
(29)~\citet{maxtedetal04-1},
(30)~\citet{pyrzasetal09-1},
(31)~\citet{parsonsetal11-2},
(32)~\citet{pyrzasetal12-1},
(33)~\citet{backhausetal12-1},
(34)~\citet{rebassa-mansergasetal12-1},
(35)~\citet{drakeetal10-1},
(36)~\citet{lawetal11-1},
(37)~\citet{almenaraetal12-1},
(38)~\citet{parsonsetal12-2},
(39)~\citet{bailey+cropper91-1},
(40)~\citet{potteretal11-1},
(41)~\citet{schwarzetal09-1},
(42)~\citet{schwopeetal11-1},
(43)~\citet{pandeletal02-1},
(44)~\citet{beuermannetal11-1},
(45)~\citet{schwopeetal02-1}.\\
Notes: Very uncertain values are followed by ``:''. It generally means that the mass was assumed and not derived. 
}

\longtab{2}{
\begin{longtable}{lccccll}
\caption{\label{tab:plan} Best fits of the orbital parameters for the currently claimed planets around eclipsing PCEBs.}\\
\hline\hline
\noalign{\smallskip}
 Name	 	&	Msin(i)      &  P            & asin(i)    & e          & Ref. & Notes   \\        
		&       [Mj]         &  [yr]         & [AU]       &            &      & \\
\noalign{\smallskip}
\hline
\endhead
\noalign{\smallskip}
\hline
\endfoot
\noalign{\smallskip}
\hline
 HW Vir c	&	14.3$\pm$1.0  & 12.7$\pm$0.2  & 4.69$\pm$0.06 & 0.40$\pm$0.10 & 1 & \\
 HW Vir d	&	30-120        & 55$\pm$15     & 12.8$\pm$0.2  & 0.05:         & 1 & *\\
 HS0705+6700 c  &	31.5$\pm$1.0    &  8.41$\pm$0.05   & 3.52	   & 0.38$\pm$0.05 & 2 & *\\
 HS2231+2441 c  &	13.94$\pm$2.20  &  15.7 	   & $\sim\,5.16$    & -	           & 3 & \\
 NSVS14256825 c	&	2.8$\pm$0.3     &  3.49$\pm$0.21   & 1.9$\pm$0.3     & 0.00$\pm$0.08       & 4 & \\
 NSVS14256825 d	&	8.0$\pm$0.8     &  6.86$\pm$0.25   & 2.9$\pm$0.6     & 0.52$\pm$0.06       & 4 & \\
 NY Vir	c	&	2.3$\pm$0.3     &  7.9	           & 3.3$\pm$0.8   & -	           & 5 & \\
 NY Vir d	&	2.5: 	        &  $>$15           & $\gappr\,5.08$  & -             & 5 & \\
 NN Ser c	&	6.91$\pm$0.54   &  15.50$\pm$0.45  & 5.38$\pm$0.20 & 0.0	   & 6 & \\
 NN Ser d	&	2.28$\pm$0.38   &  7.75$\pm$0.35   & 3.39$\pm$0.10 & 0.20$\pm$0.02 & 6 & \\
 V471 Tau c	&	46-111          &  33.2$\pm$0.2    & $\sim\,12.6-12.8$ & 0.26$\pm$0.02 & 7 & *\\
 QS Vir c  	&	9.01	        &  14.4            & $\sim\,6.32$    & 0.62          & 8 & \\
 QS Vir d	&	56.59	        &  16.99	   & $\sim\,7.15$    & 0.92          & 8 & *\\
 RR Cae	c	&	4.2$\pm$0.4     &  11.9$\pm$0.1    & 5.3$\pm$0.6   & 0             & 9 & \\
 UZ For c	&	6.3$\pm$1.5     &  16+3            & 5.9$\pm$1.4   & 0.04$\pm$0.05 & 10 & \\
 UZ For d	&	7.7$\pm$1.2     &  5.25$\pm$0.25   & 2.8$\pm$0.5   & 0.05$\pm$0.05 & 10 & \\
 HU Aqr c	&	7.1             & 9.00$\pm$0.05    & 4.30          & 0.13$\pm$0.04 & 11 & \\
 DP Leo c	&	6.05$\pm$0.47   &  28.01$\pm$2.00  & 8.19$\pm$0.39 & 0.39$\pm$0.13 & 12 & \\
\end{longtable} \textit{References}.
(1)~\citet{beuermannetal12-2},
(2)~\citet{beuermannetal12-1},
(3)~\citet{qianetal10-3},
(4)~\citet{almeidaetal12-1},
(5)~\citet{qianetal11-2},
(6)~\citet{beuermannetal10-1},
(7)~\citet{kundra+hric11-1},
(8)~\citet{almeida+jablonski11-1},
(9)~\citet{qianetal12-1},
(10)~\citet{potteretal11-1},
(11)~\citet{gozdziewskietal12-1},
(12)~\citet{beuermannetal11-1}.\\
$*$ The claimed third body is more consistent with a BD than with a planet.\\
Notes: Very uncertain values are followed by ``:''.
}

\normalsize

\begin{appendix}

\section{Notes on individual systems}

\subsection{Eclipsing sdB+MS/BD PCEBs}\label{app:sdBs}

\textit{HW\,Vir} is the prototype of HW\,Vir-like systems consisting of an sdB star with a late-type MS or BD companion. HW\,Vir systems  have short periods ($\sim\,2-3$ hr), and the light curves show very sharp primary and secondary minima and a strong reflection effect. HW\,Vir itself was discovered by \citet{menzies+marang86-1}, who determined the period. \citet{kilkennyetal94-1} observed a change in the orbital period for the first time, which motivated further studies \citep[e.g.,][]{cakirh+devlen99-1,wood+saffer99-1,kissetal00-1,kilkennyetal00-1,kilkennyetal03-1,ibanogluetal04-1}. Many possible explanations have been proposed and discussed, converging towards the existence of a third object with a long period and low mass. \citet{qianetal08-1} present new mid-eclipse times obtained from 2006 to 2008 that show some deviation from the sinusoidal fit proposed by \citet{kilkennyetal03-1} and \citet{ibanogluetal04-1}. They suggest a combination of cyclic variations plus a continuous decrease in the orbital period, which may reveal a fourth object with a long period. \citet{leeetal09-1} present 41 new mid-eclipse times taken from 2000 to 2008 and combine these with data from the literature. The Observed minus Calculated (O-C) diagram of the orbital period spanning more than 24 years shows a combination of two sinusoidal variations, probably produced by the presence of two substellar companions, plus a continuous period decrease that is too strong to be caused by gravitational radiation. Recently, \citet{beuermannetal12-2} have published 26 new mid-eclipse times obtained between 2008 February and 2012 February, which deviate significantly from the \citet{leeetal09-1} prediction. They also find that the solution presented by \citet{leeetal09-1} is unstable and propose a new solution involving two companions to HW\,Vir: a planet and a BD or low-mass star.

\textit{HS\,0705+6700} is an sdB+dM binary discovered to be a detached short-period eclipsing system by \citet{drechseletal01-1}. \citet{qianetal09-1} obtained 38 mid-eclipse times between 2006 and 2008, updated the ephemeris, and performed an O-C diagram including 31 mid-eclipse times obtained from the literature since 2000 October \citep{drechseletal01-1,niarchosetal03-1,nemethetal05-1,kruspeetal07-1}. They detected cyclic variations that were attributed to the light-travel time effect produced by the presence of a third object. \citet{qianetal10-3} present new mid-eclipse times taken in 2009, propose the existence of a continuous decrease in the orbital period, and derive a mass for the third body corresponding to a substellar object, probably a BD. \citet{camurdanetal11-1} obtained new mid-eclipse times in 2010 December, which disagree with a long-term period decrease. They used a sinusoidal fit to adjust a third object in the system, which may be substellar or a very low-mass star, depending on the inclination. Recently, \citet{beuermannetal12-1} have published new mid-eclipse times obtained between 2009 August and 2011 December, updated the ephemeris, and derive the parameters for a possible third body that seem to be more consistent with a substellar object.

\textit{HS\,2231+2441} was discovered to be an eclipsing sdB with a low-mass, probably substellar, companion by \citet{ostensenetal07-1}, who determined the ephemeris and estimated the masses. Mid-eclipse times monitored since 2006 \citep{qianetal10-3} reveal a continuous decrease and cyclic variations in the orbital period. The cyclic oscillation suggests there is a third object in the system, and \citet{qianetal10-3} estimate it may be a BD. 
 
\textit{NSVS\,14256825} was found to be an eclipsing sdB+dM binary by \citet{wilsetal07-1}, who presented 19 primary mid-eclipse times obtained between 2007 June and September and derived the ephemeris and orbital parameters. The system has also been monitored since 2006 by \citet{qianetal10-3}, whose preliminary results suggest a cyclic change in the orbital period. New mid-eclipse times obtained between 2010 September and 2011 October were published by \citet{kilkenny12-1}. They found an increase in period, but a long baseline is needed to see whether this behavior is cyclic. In a parallel work, \citet{beuermannetal12-1} present new mid-eclipse times obtained between 2009 July and 2011 October, which reveal an abrupt and continuous increase in the period after 2009. This is interpreted by the authors as the possible response to a third body in a highly elliptic orbit. In a very recent work, \citet{almeidaetal12-1} have reanalyzed the system including ten new mid-eclipse times between 2010 July and 2012 August and find that the variations observed in the O-C diagram can be explained by two circumbinary giant planets.

\textit{NY\,Vir} is an sdB+dM binary whose eclipsing nature was revealed by \citet{kilkennyetal98-1}. Subsequently, \citet{kilkennyetal00-1} present mid-eclipse times from 1996 to 1999, determined the ephemeris, and find no significant changes in the orbital period. However, \citet{kilkenny11-1} combine the previous results with new mid-eclipse times taken between 2001 and 2010 and find a continuous period decrease, which is too high to be due to gravitational radiation. The strong decrease was also seen by \citet{camurdanetal11-1}, who have presented new mid-eclipse times from 2009 and 2011 and adjusted a downward parabola to the O-C diagram. Recently, \citet{qianetal11-2} have combined the previous results with nine new mid-eclipse times obtained in 2011 May, updated the ephemeris, and propose that the O-C diagram can be adjusted by a downward parabola plus a periodic variation produced by a planet orbiting the primary. They also suggest that the continuous period decrease may be part of a cyclic variation that may indicate the presence of a fourth object. 

\textit{2M1938+4603} was found to be an eclipsing sdB+dM binary by \citet{ostensenetal10-1}, who derived the ephemeris and orbital parameters. The available mid-eclipse times only cover the period between 2008 June and 2010 May and therefore no conclusion about apparent period variations can be drawn yet. 

\textit{NSVS\,07826247} is the longest period sdB+dM binary known to be eclipsing so far, discovered by \citet{kelley+shaw07-1}. Mid-eclipse times have been published by \citet[][ 2008 February to 2009 March]{foretal10-1}, \citet[][ 2009 March to August]{zhu+qian10-1}, and \citet[][ 2011 February to October]{backhausetal12-1}. No evidence of period variation has been found so far, but more observations are needed in order to discard variations. 

\textit{BUL-SC16\,335} was found to be an eclipsing system by \citet{polubeketal07-1}, who suggest that it might be an HW\,Vir-like system based on the appearance of the light curve. They derived the ephemeris and estimate some of the orbital parameters. \citet{tello+jablonski10-1} rederived the orbital parameters and find disagreement with \citet{polubeketal07-1}. The parameters are still poorly constrained and no mid-eclipse times are available so far. 

\textit{PG\,1621+4737} has been found in the course of the MUCHFUSS \citep[massive unseen companions to hot faint underluminous stars from SDSS,][]{geieretal11-2} project by \citet{geieretal10-1}, who determined the period and estimated that the companion to the sdB star is a very late-type MS star or a BD. The system parameters are not constrained and there are still no published mid-eclipse times. 

\textit{SDSSJ0820+0008} was discovered by \citet{schaffenrothetal11-1}, who detected the short period radial velocity variations, observed the eclipses in the light curve indicating an HW\,Vir-like systems, and determined the period. The mass of the sdB primary covers a wide range of possible solutions, and the companion seems to be a BD. The ephemeris was derived by \citet{geieretal11-1}. Based on high-resolution spectroscopy with ESO-VLT/UVES \citet{geieretal11-4} confirm the substellar nature of the companion. They also find a significant shift in the system velocity with respect to the previous study, which may be produced by a third object in the system. However, more observations are needed. 

\textit{ASAS\,10232} was discovered by \citet{schaffenrothetal11-1}, who determined the period and observed the primary eclipse in the light curve. The reflection effect, typical of HW\,Vir-like systems, is clearly observed. The orbital parameters were derived based on photometry and spectroscopy, and no mid-eclipse time has been published so far. 

\textit{AA\,Dor} is a short-period eclipsing binary containing an sdOB primary \citep{kilkennyetal78-1}. Although many subsequent investigations have been published \citep[e.g.,][]{kilkennyetal79-1,kilkennyetal81-1,kudritzkietal82-1,rauch00-1,hilditchetal03-1,fleigetal08-1,vuckovicetal08-1,rucinski09-1,mulleretal10-1,klepp+rauch11-1}, the nature of the companion is still not clear, with some authors favoring a BD companion while others favor a very low-mass M star. Eclipse timings were studied by \citet{kilkenny00-1}, who found no evidence of any period variations between 1977 and 1999. \citet{kilkenny11-1} increased the baseline by including 13 new primary eclipse timings obtained between 2000 and 2010, updated the ephemeris, and confirmed the stability of the orbital period.

\textit{EC\,10246-2707} has recently been reported as an eclipsing sdB+dM binary by \citet{barlowetal12-1}, who estimate the orbital parameters, derive the ephemeris, and present 49 mid-eclipse times covering 15 years between 1997 February and 2012 June. The O-C diagram reveals no secular changes larger than $10^{-12}$ s s$^{-1}$ in the period. However, the relatively low precision of the timings does not allow to rule out small-amplitude variations such as e.g. those observed in NN\,Ser. Additional observations of eclipses with high precision are needed.

\subsection{Eclipsing WD+MS PCEBs}\label{app:WDMS}

\textit{NN\,Ser} was discovered to have deep eclipses by \citet{haefner89-1}, who classified the system as a DA+dM binary and measured the orbital period. \citet{brinkworthetal06-1} performed high-time-resolution photometry with ULTRACAM and detected a decrease in the orbital period that Applegate's mechanism fails to explain. They suggest it may be related to the presence of a third body. \citet{qianetal09-2} propose a sinusoidal fit to the O-C diagram. Later, \citet{parsonsetal10-1,parsonsetal10-2} presented the result of eclipse observations performed with ULTRACAM since 2002, which disagree with the sinusoidal fit proposed by \citet{qianetal09-2}. However, they still consider that a third body may be the cause of the observed period changes. \citet{beuermannetal10-1} monitored the system during 2010 and combined their new mid-eclipse times with all the previously published times, reanalyzing those where it was necessary. They conclude that the large amplitude variations observed in the period can only be caused by a third body, and even suggest that the best model is obtained with two giant planets around the binary. However, the existence of the fourth body is still rather uncertain. 

\textit{V471\,Tau} is a DA+dK2 system discovered to be eclipsing by \citet{nelson+young70-1}. Period variations have been observed for a long time \citep[e.g.,][]{lohsen74-1}, and many possibilities were suggested in the past to explain the shape of the O-C diagram \citep[e.g.,][]{ibanogluetal94-1,ibanogluetal05-1,guinan+ribas01-1,kaminskietal07-1}, such as perturbations by a third body in a long-period orbit, apsidal motion due to a low orbital eccentricity, or even mass transfer. Recently, \citet{kundra+hric11-1} have detected a change in the O-C diagram trend. This allows them to exclude mass transfer and other models that predict a further increase in the O-C value. After modeling the system, the authors find that the third component could be a BD or a very low-mass star, at a period of 33.2 years. 

\textit{QS\,Vir} was discovered in the Edinburgh-Cape faint blue object survey of high galactic latitudes \citep{kilkennyetal97-1}, where the eclipses revealed its binary nature. \citet{odonogueetal03-01} suggest that it is a hibernating CV, which was questioned by \citet{ribeiroetal10-1} and ruled out by \citet{parsonsetal11-1}, who confirm that it is a detached system and not a hibernating CV based on high-resolution UVES spectra. Orbital period variations were analyzed by \citet{qianetal10-2}, who combined new and previously published mid-eclipse times, and propose that there is a giant planet and a continuous decrease in period due to magnetic braking. \citet{parsonsetal10-2} update the O-C diagram by including ULTRACAM photometry and find strong disagreement with the fit of \citet{qianetal10-2}. They conclude that the decrease in orbital period is part of a cyclic variation that cannot be explained by Applegate's mechanism. A third body seems to be the most probable solution. Recently, \citet{almeida+jablonski11-1} have presented new mid-eclipse times and suggest that the best fit in the O-C diagram is obtained with a model with two circumbinary bodies, most likely a giant planet and a BD. 

\textit{RR\,Cae} is a WD+dM binary discovered as a high proper motion object by \citet{luyten55-1}. The eclipses were first announced by \citet{krzeminski84-1}, and further observations of the eclipses were presented by \citet{bruch+diaz98-1} and \citet{maxtedetal07-2}. The later paper updated the ephemeris and finds no evidence of variations in the orbital period on a long time scale ($\sim\,10$\,yr). \citet{parsonsetal10-2} performed ULTRACAM photometry for the system, obtained two new mid-eclipse times, and combined these with all the previous eclipse times available in order to study possible period variation. They obtained a roughly sinusoidal variation in the O-C diagram, which can be explained via Applegate's mechanism. \citet{qianetal12-1} have recently obtained six new mid-eclipse times that combined with those from the literature, show some evidence of a third object, a giant planet, and even possible evidence of a fourth companion. The last needs to be confirmed.

\textit{DE\,CVn} is a bright eclipsing WD+dM binary discovered as an X-ray source by ROSAT \citep{vogesetal99-1}. Recently, \citet{parsonsetal10-2} have obtained high-time-resolution photometry with ULTRACAM to obtain an accurate ephemeris. They combined their new mid-eclipse times with older times available in the literature \citep{robb+greimel97-1,vandenbesselaaretal07-1,tasetal04-1} in order to study possible period variations. However, only the ULTRACAM data are reliable, and there are still too few to analyze possible long-term period changes. 

\textit{GK\,Vir} was discovered by \citet{greenetal78-1}, who listed nine mid-eclipse times from 1975 April to 1978 February, and determined the orbital period. For more than twenty years there were no new eclipses observed. Between 2002 and 2007 \citet{parsonsetal10-2} observed seven primary eclipses with ULTRACAM and improved the ephemeris. After combining their points with those from \citet{greenetal78-1}, they observed a period increase and a slight variation in O-C times. \citet{drakeetal10-1} also published one mid-eclipse time for 2005 April. Recently, \citet{parsonsetal12-1} have obtained a new high-precision mid-eclipse time on April 2010 using ULTRACAM, which shows a clear deviation from linearity. The magnitude of the period change is small and therefore can be caused by Applegate's mechanism or due to a third body in the system. More data is needed before the true cause of this period change can be established. 

\textit{RX\,J2130.6+4710} was discovered to have eclipses by \citet{maxtedetal04-1}, who determined the ephemeris and published three ULTRACAM mid-eclipse times obtained in 2002 and 2003. No new mid-eclipse times are available. 

\textit{SDSSJ0110+1326} and \textit{SDSSJ0303+0054} were identified as eclipsing binaries by \citet{pyrzasetal09-1}, who studied the eclipses between 2006 September and 2007 October, derived the ephemeris, and listed four and seven mid-eclipse times, respectively. Subsequently, \citet{parsonsetal10-2} analyzed accurate eclipses obtained during 2007 October using ULTRACAM and listed one and three new mid-eclipse times, respectively. They found some deviations from the ephemeris derived by \citet{pyrzasetal09-1} for SDSSJ0110+1326, while the eclipses for SDSSJ0303+0054 appear to be consistent in the two studies. However, given the short baseline and the large uncertainty in the mid-eclipse times derived by \citet{pyrzasetal09-1} further accurate observations are needed before long-term period changes can be explored. 
Recently, \citet{backhausetal12-1} have presented six new mid-eclipse times for SDSSJ0303+0054 between 2011 August and November, which increases the baseline for this system. The new eclipses are still consistent with linear ephemeris, however the authors still do not exclude long-term period variations.

\textit{SDSSJ0857+0342} was first listed as an eclipsing system by \citet{drakeetal10-1}, who observed the regular eclipses as part of the Catalina Realtime Transient Survey and determined the short orbital period, which makes it the closest detached WD+dM binary. They published a mid-eclipse time for 2005 April. Recently, \citet{parsonsetal11-2} have presented nine new mid-eclipse times obtained between 2010 December and 2011 January with ULTRACAM, and updated the ephemeris that is so far consistent with the eclipse time listed in \citet{drakeetal10-1}. Independently, \citet{backhausetal12-1} obtained seven more mid-eclipse times between 2010 November and 2011 October. Due to the uncertainty in Drake's eclipse, possible period changes cannot be discarded so far. 

\textit{SDSSJ1210+3347} was discovered to be eclipsing by \citet{pyrzasetal12-1}, who obtained high-time-resolution photometry for nine eclipses between 2009 April and 2011 June using RISE on the Liverpool Telescope. They determined the orbital period and the ephemeris, however the mid-eclipse times still cover a short period of observation to study changes in the period. 

\textit{SDSSJ1212-0123} was listed as an eclipsing WD+dM binary by \citet{nebot-gomez-moranetal09-1}, who obtained six mid-eclipse times between 2007 January and 2008 May and derived the ephemeris. \citet{parsonsetal12-1} used ULTRACAM to obtained the first high-precision mid-eclipse in 2010 April, which was consistent with the mid-eclipse times from \citet{nebot-gomez-moranetal09-1}. They updated the ephemeris using the new accurate eclipse. 

\textit{SDSSJ1435+3733} is a partially eclipsing binary discovered by \citet{steinfadtetal08-1}, who observed three eclipses in 2007 May and June. In a parallel study \citet{pyrzasetal09-1} independently identified the system, observed seven eclipses between 2007 February and May, and updated the ephemeris.  

\textit{SDSSJ1548+4057} was found to be eclipsing by \citet{pyrzasetal09-1}, who derived the ephemeris and listed seven mid-eclipse times from May to 2008 July. Recently, \citet{backhausetal12-1} presented six new mid-eclipse times between 2011 May and August, which are consistent with linear ephemeris but are still not enough to exclude long-term period variations.

\textit{CSS06653}, \textit{CSS080502}, \textit{CSS21616}, \textit{CSS38094}, \textit{CSS40190}, \textit{CSS41631}, and \textit{WD\,1333+005} are all WD+dM binaries discovered to be eclipsing by \citet{drakeetal10-1} as part of the Catalina Realtime Transient Survey. They only list one mid-eclipse time in 2005 April or May. Recently,  \citet{backhausetal12-1} present new mid-eclipse times obtained between 2011 January and October for all these systems, and derived the ephemeris again. The accuracy of Drake's eclipses is not good enough to confirm or exclude long-term period variations. However, most of the systems seem to be consistent with linear ephemeris (within a 2-$\sigma$ error), except for CSS06653 and WD\,1333+005, which exhibit an increase in the observed period that is more than the expected error with respect to Drake's mid-eclipse times. More observations are necessary to confirm or discard these trends.

\textit{CSS07125}, \textit{CSS080408}, \textit{CSS09704}, \textit{CSS09797}, \textit{CSS21357}, and \textit{CSS25601} were also discovered to be eclipsing WD+dM binaries by \citet{drakeetal10-1} with one mid-eclipse time in 2005 April or May listed, except in the case of CSS21357 where no eclipse time is presented. No further studies are available for these systems. 

\textit{PTFEB11.441}, \textit{PTFEB28.235} and \textit{PTFEB28.852} are three eclipsing systems recently discovered by \citet{lawetal11-1} during the PTF/M-dwarf survey \citep{lawetal11-2} for transiting planets around M-dwarfs. Only one mid-eclipse time is available for each system in 2010 August, September, and November, respectively. 

\textit{KIC-10544976} is a WD+MS PCEB in the field of Kepler mission, recently published by \citet{almenaraetal12-1}. The authors list then mid-eclipse times: six in 2005, three in 2006, and one in 2008. Currently, the system has been continuously monitored by Kepler over two years, which makes it a good candidate to look for changes in the period in the near future.

\textit{SDSS J0821+4559}, \textit{SDSS J0927+3329}, \textit{SDSS J0946+2030}, \textit{SDSS J0957+3001}, \textit{SDSS J1021+1744}, \textit{SDSS J1028+0931}, \textit{SDSS J1057+1307}, \textit{SDSS J1223-0056}, \textit{SDSS J1307+2156}, \textit{SDSS J1408+2950}, \textit{SDSS J1411+1028} and \textit{SDSS J2235+1428} are all WD+MS PCEBs that were found to be eclipsing very recently by \citet{parsonsetal12-2} by correlating the SDSS WD+MS catalog \citep{rebassa-mansergasetal12-1} with Catalina Real-time Transient Survey light curves. Only one mid-eclipse time is available for each system.

\subsection{CVs with suspected planets}\label{app:CVs}

\textit{UZ\,For} is an eclipsing polar (or AM Her type CV) discovered by \citet{beuermannetal88-1}. The orbital period has been analyzed by several authors \citep[e.g.,][]{ramsay94-2,imamura+steiman-cameron98-1,perrymanetal01-1}, who detected variations from linearity in the O-C diagram. \citet{daietal10-1} detected an increase in the orbital period and possible cyclical changes, which they suggest may be produced by a third low-mass object in the system. Recently, \citet{potteretal11-1} have combined new high-speed photometry spanning ten years with previous mid-eclipse times to obtain a baseline of 27 years. They find that the O-C diagram is described best by a combination of two cyclic elliptical terms, probably due to the presence of two giant planets, and a secular variation. 

\textit{HU\,Aqr} is also an eclipsing polar discovered by \citet{schwopeetal93-3} during the ROSAT All Sky Survey. \citet{schwarzetal09-1} updated the ephemeris and listed 72 eclipse egress times obtained between 1993 and 2007, revealing complex deviations from a linear trend in an O-C diagram. \citet{qianetal11-1} combined these points with ten new eclipse egress times obtained between 2009 May and 2010 May, and find two cyclic variations in the O-C curve and a long-term period decrease that cannot be explained by gravitational radiation or Applegate's mechanism. They propose that the cyclic variations observed are due to the presence of two giant planets, and the long-term period decrease may reveal a third planet. However, the proposed orbital parameters for the two possible planets have been questioned by \citet{horneretal11-1}, who find the solutions extremely unstable on short time scales. \citet{wittenmyeretal11-1} used the same 82 eclipse egress times to fit different single- and double-planet models to the O-C diagram. They found that the best fits are obtained with two planets and that there is no need to invoke a third planet. Although their solutions are significantly different from the one given by \citet{qianetal11-1}, the new parameters for the possible planets are still dynamically unstable, which casts some doubt on their existence. They speculate about other mechanisms being responsible for the observed variations, such as changes in the shape of the secondary star due to dynamo effect. However, in a very recent work, \citet{gozdziewskietal12-1} publish almost 60 new eclipse egress times with better accuracy. Combining these with reanalyzed previous data, the authors find that a single circumbinary companion gives the best explanation for the O-C curve.

\textit{DP\,Leo} was the first eclipsing polar discovered \citep{biermann85-1}. A decrease in the binary period was noticed by \citet{schwopeetal02-1} and \citet{pandeletal02-1}. \citet{qianetal10-1} find a reversal of this trend, suggesting a sinusoidal variation that may be related to the presence of a giant planet. \citet{beuermannetal11-1} obtained accurate mid-eclipse times of the WD between 2009 March and 2010 February, updated the ephemeris, and combined their results with all the mid-eclipse times available by 2002, as published by \citet{schwopeetal02-1}. The data spanning more than 30 years since 1979 suggest there is a third body orbiting the binary, most likely a giant planet. 

\end{appendix}

\bibliographystyle{aa}
\bibliography{aamnem99,aabib}

\end{document}